\documentclass{article}

\usepackage{latexsym}

\usepackage[latin2]{inputenc}

\begin{document}

\title{Giant Viruses of the Kutch Desert} 

\author{Csaba Kerepesi,$^{\rm a}$, Vince Grolmusz\,$^{\rm a, b}$\footnote{to whom correspondence should be addressed}
\\
\small $^{\rm a}$ PIT Bioinformatics Group, Eötvös University,\\
 \small Pázmány Péter stny. 1/C, H-1117 Budapest, Hungary\\
\small $^{\rm b}$ Uratim Ltd.,  H-1118 Budapest, Hungary}

\maketitle
\date{}

\begin{abstract}
The Kutch desert (Great Rann of Kutch, Gujarat, India) is a unique ecosystem: in the larger part of the year it is a hot, salty desert that is flooded regularly in the Indian monsoon season. In the dry season, the crystallized salt deposits form the "white desert" in large regions. The first metagenomic analysis of the soil samples of Kutch was published in 2013, and the data was deposited in the NCBI Sequence Read Archive. At the same time, the sequences were analyzed phylogenetically for prokaryotes, especially for bacterial taxa. 

In the present work, we are searching for the DNA sequences of the recently discovered giant viruses in the soil samples of the Kutch desert. Since most giant viruses were discovered in biofilms in industrial cooling towers, ocean water and freshwater ponds, we were surprised to find their DNA sequences in the soil samples of a seasonally very hot and arid, salty environment. 
\end{abstract}

\section{Introduction} 

The discovery of new giant viruses caused a considerable turmoil in virology in the last decade: these viruses are larger than numerous bacteria and may have even more than 2,500 genes \cite{Raoult2004, Colson2011, Boyer2009, Yau2011}. They are parasitic to amoeba cells living in freshwater reservoirs or seawater habitats. Until now, they were not reported to be found in soil samples or arid environment.

The {\it Acanthamoeba polyphaga mimivirus} was first found in a cooling tower of Bradford, England in 1992, and was later identified as the first giant virus in 2003 \cite{LaScola2003}. Its genome consists of 800,000 basis pairs (bp). 

{\it Marseillevirus} was found in the biofilm of a cooling tower near Paris \cite{Boyer2009a}; its genome contains 368,000 bp.

The {\it Cafeteria roenbergensis virus (CroV)} was discovered in the seawater off the Texas coast in the early 1990s \cite{Garza1995,Fischer2010}; its genome contains 730,000 bp.

The {\it Megavirus chilensis} \cite{Arslan2011} was discovered in 2010 in a seawater sample off-coast Chile; it has a 1.2 million bp DNA that encodes 1,100 proteins.

Pandoraviruses \cite{Philippe2013} were discovered in 2013 and they have the largest genome for any viruses known. Their diameter is close to 1 $\mu$m. {\it Pandoravirus salinus} was found in seawater off-coast Chile, and has a 2.5 million bp genome that encodes around 2,500 proteins. {\it  Pandoravirus dulcis} was found in a garden pond in Latrobe University, Melbourne, Australia, has a 1.9 million bp genome. 

The Samba virus \cite{Campos2014} was found in surface water samples of the Amazon river system in Brazil. Its 1,200,000 bp long DNA encodes 938 proteins.

It is reported in \cite{Ghedin2005} that DNA strands similar to that of the Mimivirus can be found in the Sargasso sea environmental sequences database \cite{Venter2004}.

In the present work we analyze the Kutch desert metagenome  \cite{Pandit2014}, collected from soil samples with high salinity levels, for similarities with the DNA sequences of giant viruses, and found significant hits, characteristic to giant viruses, as detailed below.

\section{Results and discussion}

We have identified six short nucleotide sequences in the metagenomes of Kutch desert \cite{Pandit2014} that are characteristic to giant viruses (Table 1 and Table S2). The significant hits with 100\% query cover belong exclusively to giant viruses (with the sole exception of a draft bacterial genome in the case of the sequence U6).

Sequence U2 is the subsequence of U1, and it appeared separately from U1. We should add, that the very short nucleotide sequence U2=TGCATGAATATCAGCACCATTTTCTACCAAATATTTGAC seems to be characteristic exclusively to giant viruses if we require 100 \% query cover in a sequence search.

\begin{table}

\begin{tabular}{ | l | l | l | }
\hline
	ID & Biosample & Name of the hit \\ \hline
	U1 & S3 & Acanthamoeba polyphaga mimivirus, complete genome \\ \hline
	 & \  & \  \\ \hline
	U2 & S3 & Samba virus, partial genome \\ \hline
	 && Mimivirus terra2 genome  \  \\ \hline
	 && Hirudovirus strain Sangsue, complete genome  \  \\ \hline
	 && Acanthamoeba castellanii mamavirus strain Hal-V, complete genome  \  \\ \hline
	 && Acanthamoeba polyphaga mimivirus isolate M4, complete genome  \  \\ \hline
	 && Acanthamoeba polyphaga mimivirus, complete genome  \  \\ \hline
	 && Acanthamoeba polyphaga mimivirus, complete genome  \  \\ \hline
	 && Moumouvirus Monve isolate Mv13-mv, partial genome  \  \\ \hline
	 && Moumouvirus Monve isolate Mv13-mv, partial genome  \  \\ \hline
	 && Acanthamoeba polyphaga moumouvirus, complete genome  \  \\ \hline
	 && Moumouvirus Monve isolate Mv13-mv, partial genome  \  \\ \hline
	 && Megavirus chiliensis, complete genome  \  \\ \hline
	 && Megavirus lba isolate LBA111, complete genome  \  \\ \hline
	 && Megavirus courdo11, complete genome  \  \\ \hline
	 && Megavirus terra1 genome  \  \\ \hline
	 && Megavirus courdo7 isolate Mv13-c7, partial genome  \  \\ \hline
	 && \   \  \\ \hline
	U3 & S1 &  Pandoravirus salinus, complete genome \\ \hline
	 & \  & \  \\ \hline
	U4  & S7 & Samba virus, partial genome \\ \hline
	 && Mimivirus terra2 genome  \  \\ \hline
	 && Hirudovirus strain Sangsue, complete genome  \  \\ \hline
	 && Acanthamoeba castellanii mamavirus strain Hal-V, complete genome  \  \\ \hline
	 && Acanthamoeba polyphaga mimivirus isolate M4, complete genome  \  \\ \hline
	 && Acanthamoeba polyphaga mimivirus, complete genome  \  \\ \hline
	 && Acanthamoeba polyphaga mimivirus, complete genome  \  \\ \hline
	 && Megavirus courdo11, complete genome  \  \\ \hline
	 && Megavirus chiliensis, complete genome  \  \\ \hline
	 && Megavirus terra1 genome  \  \\ \hline
	 && Megavirus lba isolate LBA111, complete genome  \  \\ \hline
	 && Megavirus courdo7 isolate Mv13-c7, partial genome  \  \\ \hline
	 && \   \  \\ \hline
	U5 & S7 & Pandoravirus salinus, complete genome \\ \hline
	 && Pandoravirus dulcis, complete genome  \  \\ \hline
	 &&   \  \\ \hline
	U6 & S6 & Bradyrhizobium sp. Ai1a-2 K288DRAFT\_scaffold00003.3\_C,  \\ \hline
	 && Pandoravirus salinus, complete genome  \  \\ \hline
	 && Pandoravirus dulcis, complete genome  \  \\ \hline
	\  & \  & \  \\ \hline
\end{tabular}

 \caption{The short summary of the nucleotide sequence hits. A more detailed table is given as the Supporting Table S1.}
\end{table}

\section{Materials and Methods} 

The seven saline desert metagenomes \cite{Pandit2014} (see Figure S1 and Table S1 in the supporting material)  were downloaded from the Sequence Read Archive (SRA) \cite{Leinonen2011}. Next, the raw reads were converted by the NCBI SRA Toolkit (http://www.ncbi.nlm.nih.gov/books/NBK158900/) into FASTA format. 

The next step was building of seven nucleotide databases by the {\tt makeblastdb} program of the NCBI BLAST+ application suite \cite{Altschul1990} (http://www.ncbi.nlm.nih.gov/books/NBK1763/) on our local Linux server.

The seven locally built databases, containing Kutch metagenomes \cite{Pandit2014}, were searched for sequence similarities against the genomes of four giant viruses, namely: 

{\it Acanthamoeba polyphaga mimivirus}: gi$|$311977355$|$ref$|$NC\_014649.1$|$, 

{\it Megavirus chiliensis}:  gi$|$363539767$|$ref$|$NC\_016072.1$|$; 

{\it Pandoravirus dulcis}:  gi$|$526118633$|$ref$|$NC\_021858.1$|$; and 

{\it Pandoravirus salinus}:  gi$|$531034792$|$ref$|$NC\_022098.1$|$. 

The genomes of the viruses above were downloaded from the  NCBI Genome Data Base: http://www.ncbi.nlm.nih.gov/genome/. For the sequence alignments, we used the locally installed BLASTN program \cite{Altschul1990,Zhang2000,Morgulis2008}, and we have found six highly significant hits in the Kutch metagenomes (Table 1 and supporting Table S2). 

Those significant hits were searched again for evaluating specificity, using the web-interface of the NCBI BLASTN 2.2.30+ with default Megablast settings \cite{Zhang2000,Morgulis2008}, and using the Nucleotide collection (nt) database of 25,204,281 sequences in case of the first four hits. The last two hits (U5 and U6) did not produce significant alignments with Megablast, so we searched the {\tt ref seq} Genomic DB \cite{Zhang2000,Morgulis2008}.
The results of the alignments with 100 \% query cover are listed in Table 1 and with more detail in supporting Table S2.

\section{Conclusions}

We have shown, by our knowledge at the first time, the very probable presence of giant viruses in the soil of a salty, arid and very hot ecosystem, the Kutch Desert. Our result implies that not only the oceans, biofilms in cooling towers or small freshwater ponds, but desert soil can also accommodate these newly discovered viruses.


\end{document}